# Using AI to Help in the Semantic Lexical Database to Evaluate Ideas

Pedro Chávez Barrios[1], Davy Monticolo[2], Sahbi Sidhom[3]

*Abstract* — Inside a challenge of ideas there are several phases in a Creative Support System (CSS), they are problem analysis, ideation, evaluation, and implementation. Our problem: we need a full semantic lexical database SLD in an oral (voice) and writing way to help stakeholders to create ideas, these ideas contain nouns, verbs, adverbs, adjectives in the English, Spanish, and French languages. We utilize a Cloud Service Provider to use a service of Artificial Intelligence (AI), also we prepare nouns, verbs, adjectives and adverbs files in order to create the service text to voice and create our SLD with voice. This paper presents, first, an introduction about some contests that use a semantic lexical database in different languages; second, a SLD management approach using analysis of texts; third, a management application approach to complete all the new elements; fourth, the results of the management application approach, finally the conclusions and future work.

**Introduction**

Our state Querétaro and our country have some problems, similar to other countries. Our state has two principal and particular problems water and garbage management without forget many others. We have to solve these problems, our university is concerned about them, we need to use creativity and innovation to share these problems and get the solutions. Educative institutions from France, México, and USA, create challenges to produce creative ideas by mean of creative methods. We have observed that ideas could be generated in English, French, or Spanish but our semantic lexical database (SLD) is made in English according to wordnet3.0 (University, 2010), so we need to design a management tool to have a SLD in the three languages that we will need; the technical advance in México in languages makes the opportunity to analyze ideas in the English language, since this language is the most important to transmit articles, conferences and mainly ideas.

The ideas are writing in natural language so it is necessary to implement a Natural Language Processing (NLP); Our initial work is not simple, we need to analyze lexical semantic individual words, sentences, paragraphs, voice as Chowdhury express in (Chowdhury, 2005) "NLS explores how computers can be used to understand and manipulate natural language text or speech to do useful things".

The University Autonomous of Queretaro (UAQ) and the University of Lorraine (UL) at LORIA are concerned to implement an automatic evaluation of ideas, we need to take in account five phases: "problem analysis, update database, ideation, evaluation and implementation" as Gabriel explains in (Gabriel, Monticolo, Camargo, & Bourgault, 2016). Our main problem is that our semantic lexical database is not updated, many modern words are not included in the 3.0 version of the wordnet, for this reason it is essential to complete this database with some creativity support system CSS that support collaboration and use modern tools to feed our SLD.

Some creativity support systems and some cloud service providers must be used in order to get a complete database and also to add the translation of each noun, verb, adjective or adverb that helps us to complete our work. If we want to avoid ambiguity and to have a complete communication, "we need to create a support system with a complete SLD for this issue as Tutorial Points declare in (Point, 2021) examples of NLP application: Machine Translation, Fighting Spam, Information retrieval & Web search, Automatic Text Summarization, Grammar Correction, Question-answering, Sentiment analysis, and Speech engines".

*The semantic lexical database for English language*
The database used is WordNet, it has 9 files that contains data and information; there are four data-files named data.adj (adjectives), data.adv (adverbs), data.noun and data.verb, all these files named data; also, there are five index-files named index.adj (adjectives), index.adv (adverbs), index.noun, index.sense and index.verb , all these files named index (University, 2010).

The object noun will have: identifier noun, quantity of synonyms, the noun, the semantic lexical relation with other nouns, definition; we will add oral noun, oral definition (voice) in wav or mp3 format (English, Spanish and French), this will help actors of a "creative challenge of ideas such as solver participant, creative expert, technical expert, or organizer" according to the actor from creativity challenge (Chávez Barrios, Monticolo, & Sidhom, 2020) by Chávez.

[1] PhD Pedro Chavez Barrios is a profesor from Escuela de Bachilleres, Plantel Sur de la Universidad Autónoma de Querétaro; Querétaro, México; pedro,chavez@uaq.mx (**autor corresponsal**)
[2] PhD Davy Monticolo is a research professor from University of Lorraine at ERPI laboratory; Nancy, France. davy.monticolo@univ-lorraine.fr
[3] PhD Sahbi Sidhom is a research professor from University of Lorraine at LORIA laboratory (Kiwi); Nancy, France. Sahbi.sidhom@univ-lorraine.fr

The data.adj file has adjectives 18185 in his lines, we will create an object that contains the identifier adjective, quantity of synonyms, the adjective, the semantic lexical relation with other adjectives and its definition, we will add the translation in Spanish, French, drawing, voice, and actor.

Examples from Princeton University WordNet database

00002730 00 a 01 acroscopic 0 002 ;c 06076105 n 0000 ! 00002843 a 0101 | facing or on the side toward the apex
00002843 00 a 01 basiscopic 0 002 ;c 06076105 n 0000 ! 00002730 a 0101 | facing or on the side toward the base
00003552 00 s 02 emergent 0 emerging 0 003 & 00003356 a 0000 + 02631097 v 0102 + 00051513 n 0101 | coming into existence; "an emergent republic"

The Data.adv file has adverbs 3625 in his lines, the Data.adv file has adverbs 13789, there are also files in the folder dict that contain some kinds of nouns, verbs, adjectives, and adverbs like: noun_person, noun_event, noun_shape.

*The semantic lexical database for French language*

The lexique 3.8 database (Boris & Christopher, 2001) give us a 142695 words among verbs, adjectives, auxiliars, adverb, art, con, lia, pre, pro, ono, nom; All these words come from two corpus: Frantext corpus and films corpus.

*The semantic lexical database for Spanish language*

There are a Spanish database formed by two corpus CREA and CORDE but in the last twenty five years, the Real Academia Española (RAE, (Real Academia Española, 2019)) have created two new corpus "corpus del diccionario histórico de la lengua Española" and "corpus del español del siglo XXI", written by RAE (www.rae.es); unfortunately there isn't an available database with nouns, verbs, adjectives and adverbs available to researchers.

A Spanish lexical decision database from https://figshare.com/projects/SPALEX/29722 a massive online data collection gives us a database that contains data from a Spanish crowdsourced lexical decision megastudy. Aguasvivas from SPALEX project (Aguasvivas, 2018) and (Aguasvivas, Jose; Duñabeitia, 2018) collected the data through an online platform from May 12th, 2014 to December 19th, 2017. Up to that point, 209,351 participants had finished 282,576 tests by completing one (80.01%), two (14.11%), three (3.28%), or more sessions (2.60%). The majority of the data (68.88%) was acquired during the first month of the experiment, when an advertising campaign was done in order to attract the public's attention.

During a creative challenge context, actors produce knowledge by means of creative ideas that solve a specific problem. There are many lexical elements in one idea but basically, we are going to start analyzing nouns after getting the lexical semantic relation based on nouns.

The objective of this article is to propose an automatic process to complete our lexical semantic lexical database by mean of analysis of texts from Spanish, English and French languages. To achieve this proposal, our process has six steps "1- to select the actor", "2- to choose a language", "3- to get the noun, verb, adjective, adverb or other to work", "4- to capture the translation", "5.- to capture a definition", "6.- To get and image".

This article concerns in completing the semantic lexical database to help in the evaluation, order, and comparative of idea cards in a CWS; in section two, Managing Semantic lexical database approaches; in the third section, our learning managing semantic lexical database approach is presented. The last section is used for results, analysis, conclusion, and future work.

## Learning and Managing Semantic Lexical Database Approaches

*Text Analysis*

The analysis of texts is one of the most common ways to complete our semantic lexical database SLD; first, we need the words (verbs, adjectives, nouns, adverbs, etc) which will be obtained in an automatic way; second an actor have to do their translation, third we have to check the translation by a different high level actor; we will analyze free texts from the New York Times, BBC News, Le Monde, Liberation, El país, Milenio and free articles from different websites. We could select any article from each journal and every noun, article, verb, adjective must be captured in order to have a minimal quantity of the lexical database. There are several types of text analysis, Russel defines two types in (H. Russell Bernard & Gery, 1992), he says "We cover two broad types of text analysis: the linguistic tradition, which treats text as an object of analysis itself, and the sociological tradition, which treats text as a window into human experience", he analyses the indigenous literature, grounded theory, classical content analysis, semantic network analysis, cognitive mapping, and Boolean analysis. The author Tausczic (Tausczik & Pennebaker, 2009) mentions "We are in the midst of a technological revolution whereby, for the first time, researchers can link daily word use to a broad array of real-world behaviors"; nowadays, there are many computer programs, applications which are analyzing every document, so we need to create our own programs to help us to complete our SLD. In order to obtain the data, we have to identify the sources, second to create our database, store the data; also we need to identify our actors which capture an review in different levels our data; as (Humphreys & Wang, 2018) states "There are four steps to data collection: identifying, unitizing, preparing, and storing the data". Some years ago we could say, if the data is captured we could analyze it easily (Ball, 1994), but now, the enormous quantity of available data, sound and images makes the analysis a hard work even with computers.

*Photo Analysis Capture*

Today, huge quantities of images are available but unfortunately few of them are free, so we will take our own photos to capture nouns, verbs, adjectives, and adverbs related to the image. Nowadays, artificial intelligence (AI) has done a lot in languages, some examples could be the Watson assistant (IBM Cloud, 2023) (here we could use Docker (Products, 2022) to implement our translator) and to detect images, Duolingo App (DuolingoInc, 2022), and some educative translator websites like TV5MONDE (SA, 2023) and CAMBRIDGE English language at learning English (2023, n.d.).

There are many activities to create our own semantic lexical database using our own corpus; fortunately, some educative centers and institutions offer their work for free. The university of Princeton offers a lexical used in Wordnet (English language), the author of a Dataset word Information, (Aguasvivas, 2018) gives us some information (Spanish language), and Boris (French language) offers the Manuel de Lexique (Boris & Christopher, 2001). On the contrary, the Real Academia Española encrypts a lot of her information about nouns, verbs, adjectives, and adverbs but gives us two principal online corpus CREA and CORDE.

*Voice Analysis Capture*

The management of voice has some application which could help us during our learning and managing a SLD, on the web we could mention first the VoiceSauce application by Shue (Shue, Keating, & Vicenik, 2009), this application provides automated voice measurements over audio recordings and computes many voice measures, including those using corrections for formant frequencies and bandwidths; second, voice stress analysis.

*Learning and capturing*

Managing the semantic lexical database is a learning and capturing process with the purpose of getting a complete SLD. Actors will translate all the lexical database; we recommend that the actors are university students or professional people.

Our university needs a semantic lexical database and needs also to create processes to manage ideas from different languages. In the future, we will use our own cryptography process to protect our data.

We need also to find the better solutions to solve industrial problems in a creativity challenge; to get the better solutions we have to define our variables which could be defined using the Spanish, English and French grammar and create heuristics in order to get the best solution (Optimization).

**Managing Semantic Lexical Database Approach Proposal**

Our proposal to manage our Semantic Lexical Database includes adding, modifying, and deleting the principal grammar elements from the language EN, FR, and SP. The text analysis application helps us to save and organize our grammar elements in order to be used in a creativity workshop in order to select, compare and cluster ideas. We have design a process to get all the lexical semantic noun, it was designed in the poster (Barrios et al., 2018) as a model knowledge organization and as a part of a multiagent system mentioned (semantic Model Knowledge Agent) in (Chávez Barrios et al., 2020) by Chavez.

Today, there are some language that help us to feed our AI service such as python, C++, JavaScript, php, Terraform, APEX, SDKs, GraalVM, Go, Rust, .NET, Ruby but we will use C++ and Java that according to Oracle (Oracle, 2023) "Oracle Java is the #1 programming language and development platform", we will chose Java SE Platform because the complete documentation on https://docs.oracle.com/java;

With respect to some Development Environment Tools, there are some like Apache Netbean, Eclipse, IntelliJ Idea, Visual Studio Code, Tooling Platform and Application Framework; we chose Visual studio Code for its facility and Apache Netbean because is a free editor, highlights source code syntactically and semantically and easy refactor code [4]; The Apache Netbean 19 (AN) is an important tool (Copyright © 2017-2023 The Apache Software Foundation., 2023) "NetBeans IDE is a free and open source integrated development environment for developing applications on Windows, Mac, Linux, and Solaris operating systems"; Visual Studio Code uses in extensions "Extension Pack for Java" it was so easy if you have installed Java. To install AN installers and packages: Apache-NetBeans-19-bin-windows-x64.exe (SHA-512, PGP ASC) from the Apache Software Foundation Apache Download Mirrors.

---

[4] https://netbeans.apache.org/front/main/

*Using AI to get voice*

IBM Skills Build offers services that use text, voice and translator; There are some services and programs that we need to get voice from IBM like Watson AI services, Node-Red, Docker, and the service text to speech from IBM Cloud.

*Watson AI Services available on IBM Cloud*

We use the service text to speech as IBM course described in figure 3 "(\Y2024_JD\Watson) IBM" Watson Text to Speech allows users the ability to add speech transcription functionality to applications. Using Watson's AI capabilities, speech to text quickly creates accurate transcriptions for up to six different speakers in a transcript" here use the service of text to text to voice in a project on creating a voice assistant using OpenAI and IBM Watson Speech Libraries for Embed where they could use some different voices and languages.

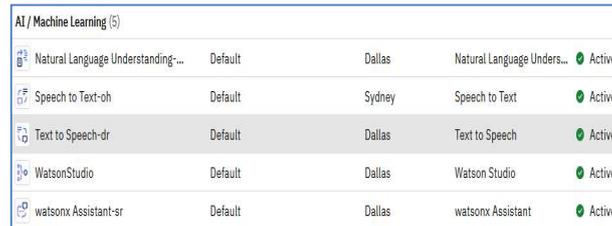

Figure 1. Active service Text to Speech from IBM Cloud

*Node Red*

Node-RED is a high-level visual programming tool, with the particularity that you don't need previous experience in the programming language. It's very easy! According to (Foundation, 2024). This tool makes use of nodes (connections of several elements), each with a specific task, you can combine the functionality of the nodes to create all kinds of programs. Discover what it is and what you can achieve with Node-RED: Open Source: Connect API (Facebook, Twitter), HW (Robots) and online services (storage, DB). Programmable devices with NR: Raspberry, PI, Arduino, Android; the Cloud; Real projects: Web pages, Chatbots, Image recognition, Controlling devices in your home, Sentiment analysis.

The curl command to get voice using AI from IBM; curl has three main parts in this curl command; in the first part of the command, the apikey is absolutely necessary to use the IBM service voice; the type of audio and the text to convert in voice.

```
curl -X POST -u "apikey:EpQH2BHH…EYk5hAzM2vnY" ^
--header "Content-Type: application/json" ^
--header "Accept: audio/wav" ^
--data "{\"text\":\" noun_definition\"}" ^
```

The second part has the name of the sound file.
```
--output noun_m.wav ^
```

The third part has type of voice in this example is a Michael voice.
```
"{https://api.us-south.text-to-speech.watson.cloud.ibm.com/instances/9521fdc7-0ed9-4d8d-9366-474cb816524e}/v1/synthesize?voice=en-US_MichaelV3Voice"
```

*Docker*

"Docker is an open source project that automates the deployment of applications within software containers" (Docker Inc, 2024); Docker Desktop enhances your development experience by offering a powerful, user-friendly platform for container management. Fully integrated with your development tools, it simplifies container deployment and accelerates your workflow efficiency. It provides a straightforward GUI (Graphical User Interface) that lets you manage your containers, applications, and images directly from your machine.

*Services*

The IBM service gives us:
- Speech to text, 500 mins per month Free.
- Text to Speech, 10,000 characters per month.
- Watsonx Assistant, 30 day trial period, up to 5,000 MAUs; We use the text to speech service.

## Results

These results are based on the WordNet database and the IBM Cloud Services. We separate the syntaxis elements in nouns, verbs, adjectives, and adverbs files similar to WordNet database. The programs were made in C++ and Node Red using services from IBM cloud.

These programs are stored in our local computer; for C++ language, the arguments are declared in Debug/ File Debug Properties, after choose Configuration Properties-Debugging where we need to write out the arguments in Command Arguments. The programs solutions give us the next results 82192 nouns and its definitions: 13789 verbs and its definitions 3625 Adverbs and its definitions 18185 Adjectives and its definitions.

*Nouns, verbs, adverbs, and adjectives format, examples*

octave| a feast day and the seven days following it.

real_time| (computer science) the time it takes for a process under computer control to occur.

burp| expel gas from the stomach; "Please don't burp at the table".

cut| with parts removed; "the drastically cut film".

anisotropically | in an anisotropic manner

*Service TextToVoice from IBM Cloud*

We use the service TextToVoice from artificial Intelligence (IBM Cloud) to get the lexical element and its voice. Example with the noun entity and its definition. The current plan lite (free) allows 10,000 characters per month.

Entity uses 39,328 bytes to get entity.wav with 6 characters.

Entity_m uses 238,998 bytes to get entitym.wav with 100 characters "that which is perceived or known or inferred to have its own distinct existence (living or nonliving)"

*Before using IBM Cloud serrvice of AI*

We need first to create our service text to voice using the IBM Cloud account and to verify if the service is active. second start the container software Docker Desktop and start our container called nombre02; third we must connect to the Node Red application and test if the service of text to voice works. Now we could use the curl command to create our voices files.

*Lexical elements per month using IBM Cloud*

We use a counter of characters to get the plan lite 10,000 characters per month and produce the command for every lexical element with curl and credentials made in a cpp program.

Parameters:

C:\Users\pichb\Documents\Y2021\EneJun\Programacion\NounsOrder01.txt

C:\Users\pichb\Documents\Y2021\EneJun\Programacion\NounsVoice01.txt

Results using command CURL, AI IBM Cloud, text to voice service and using less than 10,000 characters:

```
curl -X POST -u "apikey:my apikey_Pedro" ^
--header "Content-Type: application/json" ^
--header "Accept: audio/wav" ^
--data "{\"text\":\" noun_definition\"}" ^
--output noun_m.wav ^
"{https://api.us-south.text-to-speech.watson.cloud.ibm.com/instances/9521fdc7-0ed9-4d8d-9366-474cb816524e}/v1/synthesize?voice=en-US_MichaelV3Voice"
```

We have 7,520,838 characters in the nouns file so we will spend 7,520,838/10,000=752 months to capture, just for nouns and its definitions.

We will get 123 nouns per month on average. The total of nouns is 82,192 so we will get almost 0.14% of nouns every month. After using the curl command we have captured some nouns, we must do the same with the other lexical elements. We create 195 voice files, 15,796,783 bytes in just 7 minutes using the curl command, noun file and the noun voice file. We think that by having 30% of the nouns, verbs, adverbs and adjectives we could start the analysis of ideas.

## Conclusion

In the future, we need to use another type of AI service provider to analyze the time necessary to capture all the lexical elements, there are other cloud service providers such as Google Cloud, Alibaba Cloud, SAP, Microsoft Azure, Salesforce, Oracle Cloud and Amazon Web Services, the next article will be the comparative of the service text to voice from different cloud providers because the quantity of characters allowed by one provider is just 10,000 per month (lite or free) and we need to know more about the other cloud providers.

It is necessary to start a manual capture (voice) of the lexical elements working in a collaborative way and start the image capture and also to capture the main elements or the most used elements of our semantic lexical database. Working in collaborative way professors and students from the Universidad Autónoma de Querétaro could complete our voice semantic lexical database and also try to decrease the time used to use the 10,000 characters in 7 minutes.

**Biography of the authors**

**Dr. Pedro Chávez Barrios** is a professor at the High School of the Autonomous University of Querétaro, Mexico. He completed his doctoral studies at the Université de Lorraine, Nancy, France. He has published articles in the journals OCTA49274, SITIS2016, and Academia Journals (Chiapas, CHS064). Congresses: L'ECOLE D'ETE DU RESEAU RNI / RRI, SESSION PLÉNIÈRE (SALLE B1) and SEMINAIRE DE L'ECOLE DOCTORALE RP2E (HAL ID : HAL-01699590)

**PhD. Davy Monticolo** is a professor at the Université de Lorraine, Nancy, France and Assistant Director at ENSGSI. Creator and coordinator of "72 hours to innovate in Agile mode". He completed his doctoral studies at the Université de Lorraine, Nancy, France. He has a doctorate in Informatique and has helped many students to achieve their PhD and Post-Doc studies.

**PhD. Sahbi Sidhom** is a professor at the Université de Lorraine, Nancy, France and is affiliated with the LORIA Research Lab. Creator of book section at ISKO Maghreb 2022. His primary research interests revolve around Information Processing, Knowledge Organization, and the development of Augmented Information Retrieval Systems, particularly in the realm of Web Intelligence and Strategic Watch Applications.